\documentclass[fleqn,usenatbib]{mnras}
\usepackage{newtxtext,newtxmath}
\usepackage[T1]{fontenc}

\DeclareRobustCommand{\VAN}[3]{#2}
\let\VANthebibliography\thebibliography
\def\thebibliography{\DeclareRobustCommand{\VAN}[3]{##3}\VANthebibliography}
\usepackage{graphicx}	
\usepackage{amsmath}	

\newcommand{\msunh}{\>h^{-1}\rm M_\odot}
 
\newcommand{\Lsunhh}{\,h^{-2}\rm L_\odot}

\newcommand{\gpch}{\>h^{-1}{\rm {Gpc}}}

\usepackage{color}
\title[CSST large-scale structure analysis pipeline: I.]{CSST large-scale structure analysis pipeline: I. constructing reference mock galaxy redshift surveys}

\author[Yizhou Gu et al.]{
Yizhou Gu$^{1,2}$\thanks{E-mail: guyizhou@sjtu.edu.cn},
Xiaohu Yang$^{1,2}$\thanks{E-mail: xyang@sjtu.edu.cn},
Jiaxin Han$^{2}$,
Yirong Wang$^{2}$,
Qingyang Li$^{2}$,
Zhenlin Tan$^{2}$,
\newauthor
Wenkang Jiang$^{2}$,
Yaru Wang$^{2}$,
Jiaqi Wang$^{2}$,
Antonios Katsianis$^{3}$,
Xiaoju Xu$^{2}$,
Haojie Xu$^{2}$,
\newauthor
Wensheng Hong$^{2}$,
Houjun Mo$^{5}$,
Run Wen$^{6,7}$,
Xianzhong Zheng$^{6,7}$,
Feng Shi$^{8}$,
Pengjie Zhang$^{2,1}$,
\newauthor
Zhongxu Zhai$^{2}$,
Chengze Liu$^{2}$,
Wenting Wang$^{2}$,
Ying Zu$^{1,2}$,
Hong Guo$^{9}$,
Youcai Zhang$^{9}$,
Yi Lu$^{9}$,
\newauthor
Yi Zheng$^{3}$,
Yunkun Han$^{10}$,
Hu Zou$^{11}$,
Xin Wang$^{12,11}$,
Chengliang Wei$^{6}$,
Ming Li$^{11}$,
and Yu Luo$^{6}$
\\
$^{1}$Tsung-Dao Lee Institute, and Key Laboratory for Particle Physics, Astrophysics and Cosmology, Ministry of Education,\\
~~Shanghai Jiao Tong University, Shanghai 200240, People's Republic of China\\
$^{2}$Department of Astronomy, School of Physics and Astronomy, and Shanghai Key Laboratory for Particle Physics and Cosmology,\\ 
~~Shanghai Jiao Tong University, Shanghai 200240, People's Republic of China\\
$^{3}$School of Physics and Astronomy, Sun Yat-sen University, Zhuhai Campus, 2 Daxue Road, Xiangzhou District, Zhuhai, People's Republic of China\\
$^{4}$Shanghai Astronomical Observatory, Chinese Academy of Sciences, Nandan Road 80, Shanghai 200240, China\\
$^{5}$Department of Astronomy, University of
  Massachusetts, Amherst MA 01003-9305, USA\\
$^{6}$Purple Mountain Observatory, Chinese Academy of Sciences, 10 Yuan Hua Road, Nanjing, Jiangsu 210023, China\\
$^{7}$School of Astronomy and Space Sciences, University of Science and Technology of China, Hefei 230026, China\\
$^{8}$School of Aerospace Science And Technology, Xidian University, Xi'an 710126, People’s Republic of China\\ 
$^{9}$Key Laboratory for Research in Galaxies and Cosmology, Shanghai Astronomical Observatory; Nandan Road 80, Shanghai 200030, China\\ 
$^{10}$Yunnan Observatories, Chinese Academy of Sciences, 396 Yangfangwang, Guandu District, Kunming 650216, China\\
$^{11}$National Astronomical Observatories, Chinese Academy of Sciences, Beijing 100101, China\\
$^{12}$School of Astronomy and Space Sciences, University of Chinese Academy of Sciences (UCAS), Beijing 100049, China
}

\date{Accepted XXX. Received YYY; in original form ZZZ}

\pubyear{2023}

\begin{document}
\label{firstpage}
\pagerange{\pageref{firstpage}--\pageref{lastpage}}
\maketitle

\begin{abstract}
In this paper, we set out to construct a set of reference mock galaxy redshift surveys (MGRSs) for the future Chinese Space-station Survey Telescope (CSST) observation, where subsequent survey selection effects can be added and evaluated. This set of MGRSs is generated using the dark matter subhalos extracted from a high-resolution Jiutian $N$-body simulation of the standard   $\Lambda$CDM  cosmogony with  $\Omega_m=0.3111$, $\Omega_{\Lambda}=0.6889$,  and  $\sigma_8=0.8102$. The simulation  has a boxsize of  $1 \gpch$, and consists of  $6144^3$ particles with mass resolution $3.723 \times 10^{8} \msunh$. In order to take into account the effect of redshift evolution, we first use all 128 snapshots in the Jiutian simulation to generate a light-cone halo/subhalo catalog. Next, galaxy luminosities are assigned to the main and subhalo populations using the subhalo abundance matching (SHAM) method with the DESI $z$-band luminosity functions at different redshifts. 
Multi-band photometries, as well as images, are then assigned to each mock galaxy using a 3-dimensional parameter space nearest neighbor sampling of the DESI LS observational galaxies and groups.  Finally, the CSST and DESI LS survey geometry and magnitude limit cuts are applied to generate the required MGRSs. As we have checked, 
this set of MGRSs can generally reproduce the observed galaxy luminosity/mass functions within 0.1 dex for galaxies with $L > 10^8L_\odot$ (or $M_* > 10^{8.5} M_\odot$) and within 1-$\sigma$ level for galaxies with $L < 10^8L_\odot$ (or $M_* < 10^{8.5} M_\odot$).
Together with the CSST slitless spectra and redshifts for our DESI LS seed galaxies that are under construction, we will set out to test various slitless observational selection effects in subsequent probes. 
\end{abstract}



\section{Introduction}

Galaxies are thought to be formed and located within dark matter halos. The distribution of galaxies thus contains important information about the large-scale structure of the matter distribution. Over the past few decades, large galaxy redshift surveys have revolutionized our understanding of the cosmos by providing comprehensive maps of the distribution of galaxies across a large range of cosmic scales. Pioneering surveys like the Two-Degree Field Galaxy Redshift Survey \citep[2dFGRS;][]{2001MNRAS.328.1039C}  and the Sloan Digital Sky Survey \citep[SDSS;][]{2000AJ....120.1579Y} have been instrumental in charting the positions and redshifts of millions of galaxies, enabling precise measurements of their clustering patterns. These surveys, complemented by more recent endeavors such as the Dark Energy Survey \citep[DES;][]{2018ApJS..239...18A}, the Extended Baryon Oscillation Spectroscopic Survey \citep[eBOSS;][]{2016AJ....151...44D}  and the ongoing Dark Energy Spectroscopic Instrument \citep[DESI;][]{2016arXiv161100036D}, have significantly advanced our understanding of cosmic structure and evolution.

Through the statistical analysis of galaxy clustering in these surveys on both large and small scales, researchers have been able to extract crucial cosmological parameters, refine models of dark matter and dark energy, and unveil the imprint of primordial fluctuations that shaped the large-scale structure of the universe \citep[e.g.,][]{2003MNRAS.346...78H, 2004MNRAS.350.1153Y, 2007MNRAS.381.1053P, 2018ApJ...861..137S, 2021MNRAS.501.1481H, 2021PhRvD.103h3533A, 2022PhRvD.105b3520A, 2023arXiv231003066X}. 
In addition to these cosmological probes, the clustering of galaxies, especially on small scales has also been used to infer the galaxy-halo connections, which significantly enhanced our understanding of the galaxy formation processes \citep[e.g.,][]{1998ApJ...494....1J, 2002ApJ...575..587B, 2003MNRAS.339.1057Y, 2006ApJ...647..201C, 2009MNRAS.392.1080S, 2012ApJ...752...41Y, 2012ApJ...744..159L, 2015ApJ...799..130R, 2016MNRAS.460.1173R, 2018MNRAS.481.5470X, 2022ApJ...928...10G, 2022MNRAS.509.3119Z, 2022ApJ...933....9L,1980ApJ...236..351D, 2023ApJ...944..200X}.

In addition to those ground-based large galaxy redshift surveys, there are a number of space endeavors that pursue the large-scale structure (LSS) studies of galaxies. Among these efforts, the EUCLID aiming at observing galaxies in 1/3 of the sky has recently obtained its very first image \citep{2011arXiv1110.3193L}. The Roman telescope aiming at observing galaxies in a redder wavelength with smaller sky coverage is supposed to launch in 2027 \citep{2019arXiv190205569A}. The Chinese Space-station Survey Telescope (CSST) is planned to conduct galaxy observations covering $\sim 17,500 \deg^2$, which would provide unprecedented data for exploring the Universe \citep{2011SSPMA..41.1441Z, 2018MNRAS.480.2178C, 2019ApJ...883..203G, ZhanHu2021...CSB}. All these endeavors have proposed both photometric imaging and slitless spectroscopic observations. However, unlike the ground fiber-fed spectroscopic redshift surveys, the slitless spectroscopic observations suffer more significantly from various contamination and incompleteness, especially caused by the overlapping of spectra along the dispersed directions. These selection effects will significantly impact the accurate measurement and interpretation of galaxy clustering.

Mock galaxy redshift surveys (MGRSs) thus play a pivotal role in understanding and interpreting observed galaxy clustering patterns, with the existence of selection effects and observational errors. These synthetic surveys, constructed from large N-body simulations, mimic the distribution of galaxies in the universe based on theoretical models and input cosmological parameters. By adding various kinds of selection effects into the MGRSs, we will be able to assess on what level the clustering patterns are affected by observations and on what level can we recover the true input values both in cosmology and galaxy formation framework \citep[e.g.][]{2004MNRAS.350.1153Y, 2012ApJ...756..127G, 2012MNRAS.424..564R, 2015MNRAS.451..660E, 2016A&A...585A.116M, 2019MNRAS.484.1285S, 2020MNRAS.498..128M, 2023MNRAS.519.1132M}. 
In addition to these, 
by comparing the statistical properties of mock surveys to future real observational data, one can validate the accuracy of their analysis techniques, test the reliability of various statistical measurements, and make more accurate model constraints \citep[e.g.][]{2017MNRAS.464.1168R, 2020MNRAS.499...89W, 2022MNRAS.516...57Y}. 
Via emulator,  such kind of efforts have also been applied to the pre-researches for the Roman mission in recent works \citep[e.g.,][]{2019ApJ...874...95Z, 2021MNRAS.501.3490Z, 2021MNRAS.505.2784Z}.

In this work, we set out to construct a set of large  MGRSs that can be obtained by an ideal spectroscopic survey,  with redshift $0<z<1$ and a magnitude limit $m_z<21$. This work serves as the benchmark for subsequent studies of various CSST observation selection effects. Here we first make use of the large Jiutian N-body simulation suite by constructing the light-cone catalogs of halo/subhalo for our study. 
Then, a $z$-band galaxy luminosity is assigned to each subhalo using the subhalo abundance matching method (SHAM) to reproduce the observed luminosity functions (LFs) based on DESI One-percent survey (DESI 1\%\footnote{\url{https://data.desi.lbl.gov/doc/glossary/\#sv3}}). 
Each galaxy in the light-cone catalogs will be matched to the closest galaxy in the DESI observation in the 3-D parameter space of redshift, luminosity, and dark halo mass from the group catalog. The application of 3-D parameter space samplings enables the provision of other photometric properties besides luminosity assigned by SHAM and allows the use of multi-band images.  
After implementing the survey geometry and magnitude limit cuts for CSST observation, as well as the foreground mask, the reference MGRSs and their group catalogs are constructed.  
Finally, we present some measurements of the MGRS with the set of spectroscopic redshifts, including the luminosity functions (LFs), stellar mass functions (SMFs), and conditional luminosity functions (CLFs) for galaxies, to demonstrate the consistency and difference between the MGRSs and DESI observations. 

The layout of the paper is organized as follows. 
In Section~\ref{sec_data}, the Jiutian simulation and DESI observation data are described. 
In Section~\ref{sec_populate}, we briefly describe the methodology of building light-cone and populating galaxies. 
In Section~\ref{sec_geo}, the survey geometry and magnitude cuts are applied. 
Section~\ref{sec_test} provides some tests on the mock catalog, compared with the measurements from the DESI galaxy catalog. 
Our conclusion and future outlook are summarized in Section~\ref{sec_conclusion}.

\section{The simulation and observation data} \label{sec_data}

Our CSST MGRS pipeline has a number of options that one can choose from, including different simulation input halo and subhalo catalogs, different (band) observational data to be reproduced, etc. (Yang et al. 2024 in preparation). Here we focus on our fiducial choices in the pipeline. 

\begin{figure*}
\centering
\includegraphics[width=1.\textwidth]{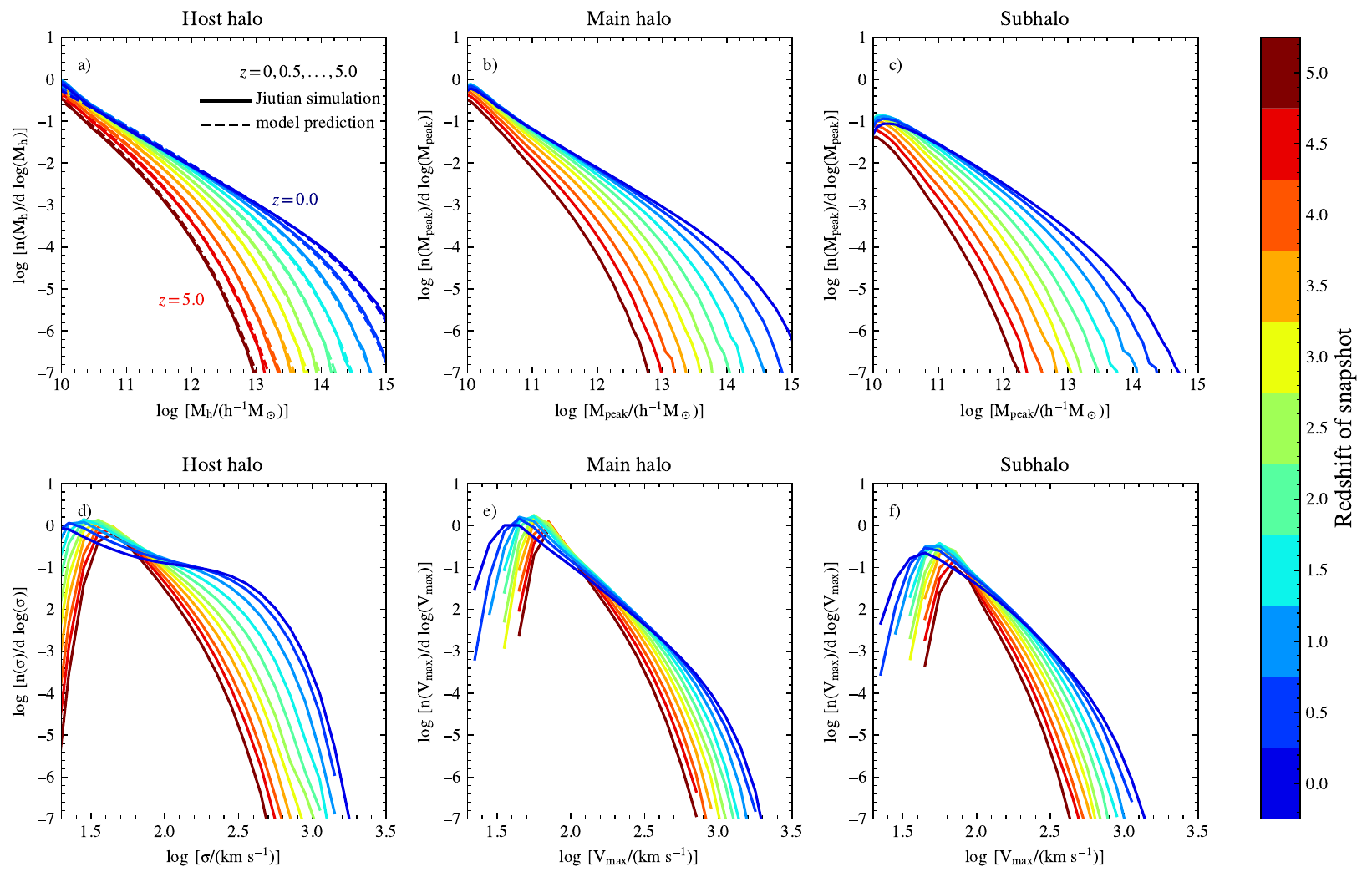}
\caption{Upper panels: the mass functions of host halo and the peak mass functions of both main halo and subhalo; The dashed lines are the halo mass functions predicted using the \protect\cite{1999MNRAS.308..119S} model for the Planck-2018 cosmology at different redshifts.  Bottom panels: the velocity dispersion functions and the maximum circular velocity functions, similar to the upper panels. 
}
\label{fig:shmf}
\end{figure*}

\subsection{Jiutian Simulations}
\label{sec:simulations} 
The fiducial simulation we use to construct our mock catalog is one high-resolution $N$-body simulation from the Jiutian simulation suite. The Jiutian suite is a series of $N$-body simulations designed to meet the science requirement for the CSST optical surveys \citep{2011SSPMA..41.1441Z, 2018MNRAS.480.2178C}. It consists of four subsets of simulations: main runs targeting the concordance cosmology, extension runs with various non-standard cosmologies, emulator runs covering a wide cosmological parameter space, and constrained runs aiming to reproduce the actual observed large-scale structure. The main runs contain three high-resolution dark-matter-only simulations run under the Planck-2018 cosmology~\citep{2020A&A...641A...6P}, with $\Omega_{M}=0.3111$, $\Omega_\Lambda=0.6889$, $\Omega_b=0.0490$, $\sigma_8=0.8102$ and $n_s=0.9665$. These simulations are run with $6144^3$ particles each, in three different periodic boxes of $0.3$, $1$, and 2 $\gpch$ per side respectively. The simulation used in this work is the 1 $\gpch$ run, with a particle mass of $m_p=3.723\times10^8 \msunh$, run with the \textsc{Gadget}-3 code~\citep{2001NewA....6...79S, 2005MNRAS.364.1105S}. The simulation starts at an initial redshift of $z = 127$ and outputs 128 snapshots to $z=0$. 


Dark matter halos are identified with the Friends-of-Friends (FOF) algorithm~\citep{1985ApJ...292..371D} with a linking length of $0.2$ times the mean inter-particle separation. These halos are further processed with the new implementation of the Hierarchical Bound-Tracing code  ~\citep[\textsc{HBT+;}][]{2012MNRAS.427.2437H, 2018MNRAS.474..604H}\footnote{\url{https://github.com/Kambrian/HBTplus}} to identify subhalos and their evolution histories. \textsc{HBT+} is a time-domain subhalo finder and tree builder that works by tracking the evolution of each halo throughout the simulation. It identifies the descendant of a halo at every snapshot as either an individual halo or a subhalo. The minimum number of particles in subhalo is set to 20. The corresponding minimum halo mass is estimated to be approximately $7.5 \times 10^9 \msunh$. 
When a subhalo is no longer resolved, it keeps track of its most-bound particle. As a result, it produces high-quality subhalo catalogs that are free from many common pitfalls associated with halo-finding~\citep[e.g.,][]{2011MNRAS.410.2617M, 2013MNRAS.436..150S, 2015MNRAS.454.3020B}. Additionally, it provides a robust and physically consistent merger tree. 

In literature, there have been quite a number of different halo and subhalo properties being used to generate galaxies with different properties \citep[][and reference herein]{2013ApJ...771...30R}. Here we focus on two sets of data that are typically used in recent years: the subhalo mass and the maximum circular velocity of the subhalo. It is expected that galaxy properties should be strongly correlated with their mass before the infall (or strip) occurs.  For the subhalos which are supposed to be associated with satellite galaxies, we use the peak mass over its entire history determined by the HBT+ code for our studies \citep{2018MNRAS.474..604H}.  
The reason is that, in comparison to the mass of surviving subhalos, satellite galaxies overall suffer less from the disruption effect than the subhalos. 

In the upper panels of Fig.~\ref{fig:shmf}, we show 
the host halo mass functions (left panel), peak mass function of main halo (middle panel), and peak mass function of subbhalo (right panel) for the Jiutian simulation at different redshifts as specified in the left panel. The velocity dispersion functions of host halo (left), the maximum circular velocity functions (middle) of main halo and subhalo (right) are given in the bottom panels of  Fig.~\ref{fig:shmf}. 
For comparison, we also plot in panel a) using dashed lines the analytical halo mass functions predicted by \cite{1999MNRAS.308..119S}. Overall, the host halo mass functions agree with these theoretical predictions quite well in all redshift ranges. 
The peak mass function of the main halo is calculated using the most massive subhalo at a given host halo. 

\begin{figure*}
\centering
  \includegraphics[width=1.\textwidth]{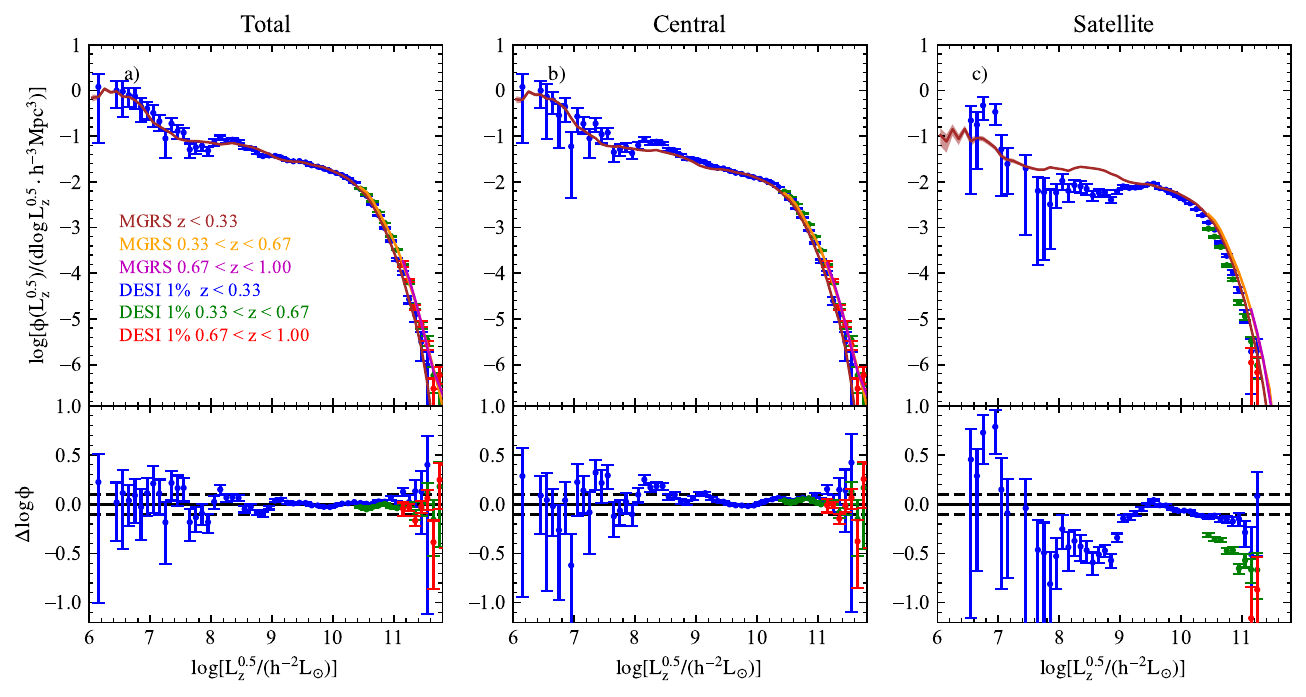} 
  \caption{Top: The total, central, and satellite luminosity functions of DESI 1\% and mock catalog in three redshift bins. The errors are estimated from the bootstrap method with 100 times resampling, denoted by the error bars for observation and shallow region for mock catalog.  
  Bottom: the difference in luminosity functions between the mock catalog and the 1\% DESI observation. The black solid lines ($\Delta = 0$) and dashed lines ($\Delta = \pm 0.1$) serve as reference lines.
  }
  \label{fig:lf}
\end{figure*} 

\subsection{DESI observational data}
\label{sec:DESI-obs}
For our fiducial observational data, we make use of the currently largest photometric redshift survey, DESI Legacy Imaging Surveys Data Release 9 (LS DR9), and DESI 1\% survey to guide the MGRS construction. The DESI LS DR9 provides the imaging in $g$/$r$/$z$ bands with 5$\sigma$ depth of 24.7/23.9/23.0 (\citealt{2019AJ....157..168D} and references therein). It consists of three independent optical surveys: the Beijing-Arizona Sky Survey (BASS), the Mayall $z$-band Legacy Survey (MzLS), and the DECam Legacy Survey (DECaLS). It is worth mentioning that although the raw data of LS were obtained from different sets of photometric systems (BASS/MzLS and DECals), the data provided by LS DR9 were reprocessed using the same pipeline and calibrated to be consistent with each other \citep[see][for details]{2019AJ....157..168D}. The data release also includes near-infrared data from the 6-year imaging of WISE, with 5$\sigma$ depth of 20.7 and 20.0 in the 3.4 and 4.6 $\mu m$ WISE bands (Wide-field Infrared Survey Explorer W1 and W2, see \citealt{2010AJ....140.1868W}). DESI 1\% survey is the spectroscopic survey, which is part of the DESI Early Data Release \citep{2023arXiv230606308D}. DESI 1\% survey contains 20 non-overlapping subfields, also referred to as ``rosettes'' \footnote{\url{https://data.desi.lbl.gov/doc/releases/edr/\#coverage-area}},  covering $\sim$ 175 $\rm deg^2$ in total. Each subfield was observed at least 12 times to ensure a fiber assignment completeness of at least 95\%.   

The galaxy catalog of $m_{z} < 21$ and $0.0 < z < 1.0$ is cut out from the observational data of LS DR9. The redshift of each galaxy is taken from the random-forest-algorithm-based photometric redshift estimation from \cite{2021MNRAS.501.3309Z}, with a typical redshift error of $\sigma_z/(1+z) \sim 0.02$. To ensure the redshift information is as accurate as possible, a small fraction of the redshifts have been replaced by the spectroscopic redshifts taken from other surveys. The galaxy catalog has removed the area within galactic latitude $|b| \leq 25^\circ$ to avoid the regions of higher stellar density.  In addition, galaxies with the PSF morphologies and close to the sources of contamination (bright star, large galaxy, and globular cluster) are also removed \citep[see more detail in][]{2021ApJ...909..143Y}. 

The extended halo-based group finder developed by \citet{2005MNRAS.356.1293Y} is applied to the galaxy catalog from imaging Surveys \citep{2021ApJ...909..143Y}. Every galaxy is assigned to a unique group and is identified as central or satellite. 
The group catalog was originally constructed using the DESI Legacy Survey Data Release 8. In this work, we have updated it to LS DR9 with additional $\sim 3 \%$ spectroscopic redshifts incorporated. The group catalog totally contains  $\sim 100$ million groups with $\sim 120$ million galaxy members having five-band photometries ($g$, $r$, $z$, $W1$, $W2$), with a sky coverage of $\sim 18200$ deg$^{2}$. In terms of photometry, LS DR9 does not incorporate many new observations. Instead, the main improvement is the reduction techniques and procedures. 


The galaxy LFs we utilize to construct the MGRSs are very close to the LFs obtained in the recent study by \cite{2312.17459v1}, which are mainly based on the DESI BGS-BRIGHT galaxies \citep{Hahn2023}. 
In that work, the measurements of LFs are divided into three redshift intervals ($\Delta z = 0.2$) up to $z = 0.6$, and the K-corrections \citep{2007AJ....133..734B} are applied to shift the redshifts to 0.1, 0.3, and 0.5, respectively. Compared with the measurements in \cite{2312.17459v1}, the primary difference in our measurements up to $z = 1.0$ is that we adopt an average K-correction band-shifted to the redshift of $z = 0.5$. The analytic function of the average K-correction of z-band we adopt is taken from \cite{2021ApJ...909..143Y}, which can be described by: $K^{0.5}_{\rm z}(z)=0.73z^2-0.54z-0.33$. Similarly, the average K-correction of the r-band is given by the formula: $K^{0.5}_{\rm r}(z)=2.01z^2-0.36z-0.74$. According to \cite{2018ApJS..236...47W}, the Sun’s absolute magnitudes are 4.61 in the r-band and 4.5 in the z-band.  

Following \cite{2312.17459v1}, 19 rosettes out of 20 in the DESI One-Percent Survey are used to calculate LFs. The excluded one is centered in the Coma Cluster, which could potentially introduce bias into the LFs. As the total number of overlapping tiles decreases in these regions, the outer region of $r_{\rm subfield} > 1.45$ $\rm deg$ of each subfield is also cut out, where $r_{\rm subfield}$ is defined as the radius from the center of each rosette. As a result, the magnitude-limited sample with $m_z < 19$ (or $m_r < 19.5$) has more than 95.0\% spectroscopic completeness. Henceforth, this sample will be denoted as ``DESI 1\%''. When calculating LFs, each galaxy is assigned a weight to take into account the Malmquist bias using the standard $V_{\rm max}$ approach. The redshift incompleteness is corrected by the application of the magnitude- and sector-dependent up-weighting. 
In our case, the redshifts of most galaxies are spectroscopic redshifts, which are considered to be reliable. Though a very small portion of galaxies only have photometric redshifts, they are given through photometric data from the full five bands ($g$, $r$, $z$, $W1$, $W2$). These galaxies with photometric redshifts only are not used for LF measurements, but are used in the group finding. Within each sector defined by Healpix \citep{2005ApJ...622..759G}, the ratio of the total number of galaxies to the number of galaxies with reliable redshifts as a function of apparent magnitude is calculated for up-weighting. Due to the high completeness of DESI 1\%, the incompleteness correction has a minimal impact on the results. 
The galaxy LFs of DESI 1\% in a few redshift bins are used to assign the luminosity of galaxies via the subhalo abundance matching (SHAM) technique (see Section~\ref{sec_sham} for more details) in this work.  

As an illustration, in Figure~\ref{fig:lf}, we show the total, central and satellite LFs of DESI 1\% color-coded by three redshift bins, $0.0<z<0.33$, $0.33<z<0.67$ and $0.67<z<1.0$, respectively. Here we do not see much evolution in the LFs as a function of redshift. 


\section{Populate subhalos with galaxies} \label{sec_populate}

In this section, we describe the details of the algorithms for populating subhalos with mock galaxies containing different observational properties.  

\subsection{Constructing subhalo light-cone catalogs}
\label{sec:light-cone}

To properly take into account the structure evolution effect in our MGRS, it's necessary for us to consider subhalo evolution at different redshifts. In our pipelines, we have two input subhalo options: (1) subhalo catalogs outputed at different snapshots/redshifts; (2) the constructed subhalo light-cone catalog.  

In option (1), we only use subhalos at different snapshots, the related light-cones are built in the following steps. Firstly, an observer is placed at a reference location in the origin box, and the boxes are replicated periodically to fill out the entire space \citep{2005MNRAS.360..159B}. The shell-like subhalo light-cone catalog is directly assembled from the snapshots, with each shell being spliced together in a sequential manner \citep[e.g.,][]{2022ApJ...928....1W}. The properties of the subhalo in the related snapshot are then used to produce a light-cone catalog of subhalo. In this case, we do not perform any interpolation of galaxy positions or other properties.  Here we have used the halo IDs to avoid repeated halos (galaxies) at the interfaces of the joining snapshots in shells \citep[e.g.,][]{Smith2022}. 

In option (2), we can use the subhalo light-cone catalog that has been constructed.  
The interpolated subhalo light-cone catalog can be pre-built. 
For example, the trace of each subhalo can be reconstructed according to the position and velocity at different snapshots \citep[e.g.,][]{2013MNRAS.429..556M}. 
The specific time at which the subhalo will be observed can be pinpointed by solving a light-cone equation. The subhalo properties at this specific time are then calculated. 
In this way, the properties of the subhalo evolve smoothly over time and are more closely aligned with their actual values, regardless of their spatial location.  


In this work, the subhalo light-cone catalog is constructed using option (1). 
The Jiutian simulation was evolved from z = 127 to z = 0 with 128 snapshots. To make spherical shells, we use 46 snapshots from No.~127 (z = 0) to No.~82 (z = 1.03). 
Meanwhile, the mock catalog, which utilizes the cubic interpolated subhalo light-cone catalog mentioned above, will be presented soon under option (2). 

%

\subsection{Subhalo abundance matching} \label{sec_sham}

To construct our mock galaxy catalogs, there are different options in our pipeline, e.g., using a CLF/CSMF modeling, the SHAM technique, and the extended SHAM, etc. (Yang et al. 2024, in preparation). Here we use the standard SHAM method to assign galaxy properties in their respective subhalos. 

SHAM is an empirical methodology employed to establish a connection between the properties of galaxies and their corresponding halos \citep{2004ApJ...609...35K, 2004MNRAS.353..189V, 2006ApJ...647..201C, 2012ApJ...752...41Y, 2018ARA&A..56..435W}. This is achieved by postulating a correlation between galaxy and halo properties, with an inherent degree of scatter. Utilizing this approach, the number density and luminosity function (or alternatively, the stellar mass function) of a specific galaxy sample can be accurately reproduced. 

In this paper, we use the $z$-band cumulative galaxy luminosity functions measured from DESI 1\% and the cumulative subhalo mass functions measured from the peak mass of all the subhalos, including central and satellite ones to link galaxies with dark matter subhalos. Here we use 6 redshift bins to calculate the cumulative LFs, and then use interpolation to obtain the cumulative LF at the subhalo redshift for our abundance matching. Note that because of the magnitude cut, the LFs at the faint end especially at higher redshifts are truncated. To properly account for those missing fainter galaxies, we have extrapolated the LFs in higher redshift bins using the ones measured in the successive lower redshift bins. 
To reduce the cosmic variance, we use all the subhalos in the closest snapshot to calculate the cumulative subhalo mass functions. 
A tentative $z$-band luminosity is assigned to each galaxy after the 
subhalo mass v.s. luminosity abundance matching. In order to take into account the scatter in the luminosity-subhalo mass relation, a scatter in the $z$-band luminosity, $\sigma_{\log(L_z)} = 0.15$ dex \citep[e.g.,][]{2008ApJ...676..248Y}, is added to each galaxy. 

\begin{figure*}
\centering
  \includegraphics[width=1.\textwidth]{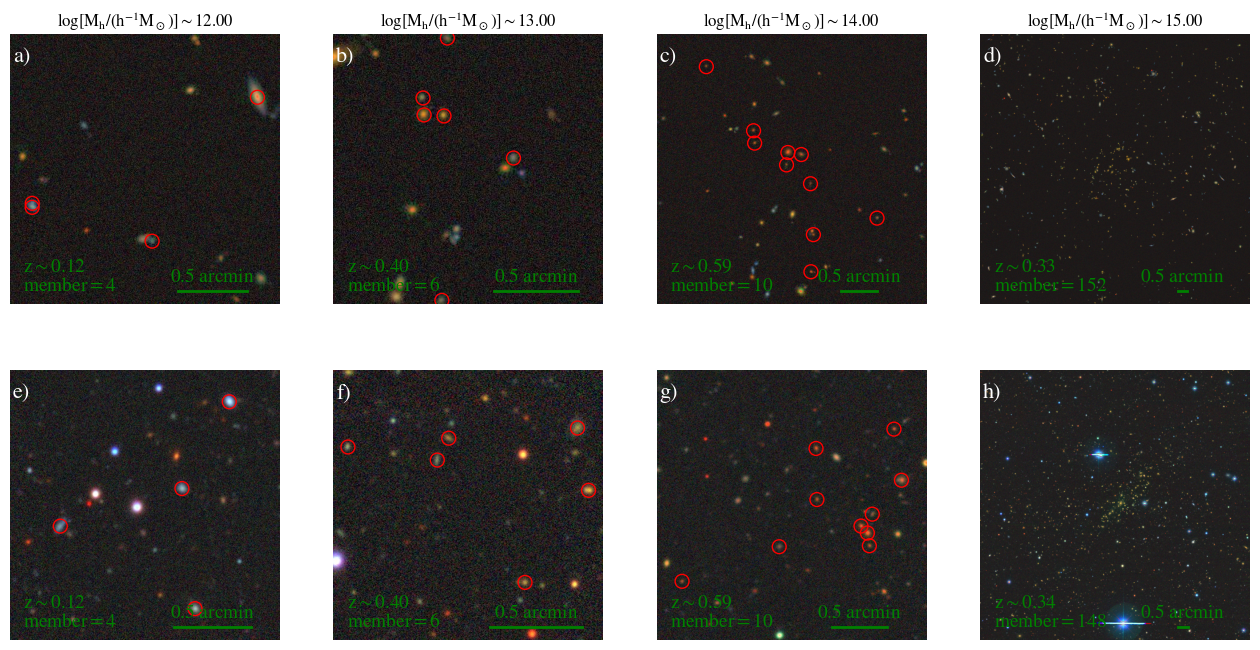}
  \caption{The comparison between the mimic images of halo in the mock catalog (upper) and the observed images of DESI clusters (bottom). The halo masses are $10^{12}$, $10^{13}$, $10^{14}$, and $10^{15}$ $\rm h^{-1} M_\odot $ from left to right. We mark the member galaxies with red circles except for the panels d) and h) due to the too much overlap. }
  \label{fig:imaging}
\end{figure*}

We show in Figure~\ref{fig:lf} the $z$-band galaxy LFs measured from our MGRS for all, central and satellite galaxies using shallow regions in three redshift bins, $0.0<z<0.33$, $0.33<z<0.67$ and $0.67<z<1.0$, respectively. Here central and satellite galaxies are those associated with main halos and subhalos. 
Our MGRS can generally reproduce the observed galaxy luminosity functions within 0.1 dex for galaxies with $\rm L^{0.5}_z > 10^8L_\odot$ and within 1-$\sigma$ level for galaxies with $\rm L^{0.5}_z < 10^8L_\odot$. 
The nearly perfect agreement between the LFs for all galaxies in observation and MGRS is rather expected because of the nature of SHAM. 
However, the LFs for central and satellite galaxies are not guaranteed. Compared to the LFs for central and satellite galaxies obtained from the DESI 1\%, decomposed by the halo-based group finder, we see they show overall good agreements. Only at the faint end with luminosity $L_z\la 10^9\Lsunhh$, the satellite galaxies tend to exhibit some discrepancies. 
This discrepancy can be induced by different galaxy-halo connections between the MGRSs and DESI observations, or the contamination when group finding. 
The overall agreement indicates that the peak mass of all the subhalos is indeed a fair tracer for the galaxy properties of the bright galaxies. 
The agreement indicates that central and satellite galaxies follow almost the same luminosity-peak mass relation for bright galaxies, which is expected if the central luminosity-peak mass relation evolves weakly over redshift~\citep[e.g.,][]{WJ10}.
Regarding the fainter galaxies, their properties might be affected by the more complex physics related to the fact of being satellites. We will come back to this point in a subsequent study with a modified SHAM model (Xu et al. 2024, in preparation). 

\subsection{Assigning galaxy properties using 3-D parameter space samplings}
\label{sec:properties} 

As the main purpose of this work is to provide a reference MGRS with sufficient information that can be used to test the selection effects of CSST slitless spectroscopic observations, we use galaxies in the DESI LS observations as our seed catalog which contain various kinds of properties including the multi-band photometries, stellar masses, SFRs, the shapes and images, and part of spectra if available, etc.  

Recent studies have shown that galaxy properties, e.g. colors and quenching fractions are mainly correlated with the intrinsic properties, e.g., SFRs, stellar mass, luminosities, and external environments, e.g., the halo mass of galaxies \citep[e.g.][]{2006MNRAS.373..469B, 2010ApJ...721..193P, 2018ApJ...852...31W, Katsianis2023}, and will evolve with redshift. 
In order to make a direct link between the mock galaxies and the observed galaxies, we match them in a 3-D parameter space using the following three essential parameters.  
The first parameter is redshift, which is necessary due to the evolving galaxy properties as cosmic time. The second parameter is z-band luminosity, which is chosen because luminosity (or stellar mass) is found to be strongly correlated with other galaxy properties, providing a link with intrinsic physics. The third parameter is halo mass, used as a tracer of the external environment.
Here, the halo mass in the LS DR9 is obtained by \citet{2021ApJ...909..143Y} using an adaptive halo-based group finder. 
It is shown that both central and satellite galaxies exhibit similar correlations of quenching efficiency with halo mass and stellar mass \citep{2018ApJ...852...31W}. This suggests that they have undergone similar quenching processes within their host halo. Therefore, we do not make a distinction between central and satellite galaxies when assigning galaxy properties. 
We employ the nearest neighbor matching between the galaxies in the MGRS and LS DR9 in the 3-D parameter space, $z|\log L_z|\log M_h$. 
To properly sample the 3-D parameter space, for luminosity and halo mass, we use their log values divided by a factor of 4 to make the match. Note that even without this rescaling, the results are not significantly impacted. 
In this way, every mock galaxy is matched with one galaxy in the observations so that we can assign the properties of observed galaxies to the mock galaxies.  

\begin{figure*}
\centering
\includegraphics[scale=0.8]{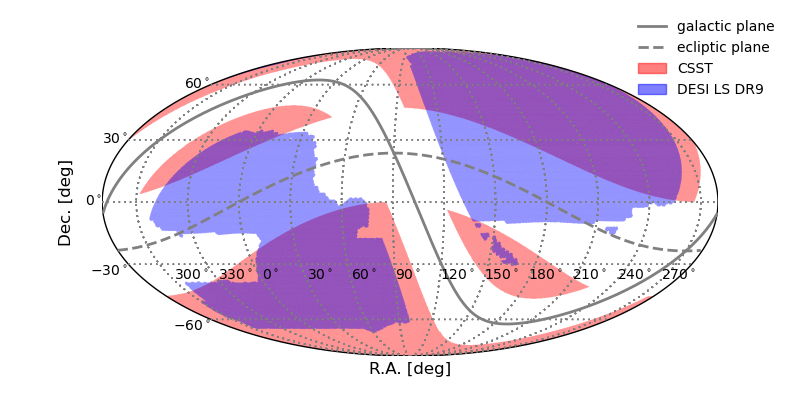} 
\caption{The sky coverages of CSST optical survey (red) and DESI LS DR9 (blue). It is shown as the Mollweide projection in celestial coordinates, with the gray solid/dashed line representing the galactic/ecliptic plane. }
\label{fig:footprint} 
\end{figure*}

Going beyond the derived galaxy properties, the images of LS DR9 can be also assigned to the mock galaxies. To illustrate this, we construct the mimic images of the halos in the mock catalog following the two main steps described below. 
\begin{itemize}
    \item Build source image library: The first step is to generate a source image library. Based on the celestial coordinates of each galaxy in the LS DR9,  we download the image cutout from the LS data server,  where the pixel size is 0.262$^{\prime\prime}$. Using the Photutils package \citep{larry_bradley_2023_7946442}, we separate the pixels of each cutout into three distinct sets: the primary object, other sources, and the background. We store the primary object linked with the related galaxy to build the source image library. The background image around each galaxy is also kept in the library. 
    \item Make the mock image: Next, according to the mock galaxy catalog, we start from galaxies with the highest to lowest redshift to insert object images. For a given sky region, we construct the blank image with the World Coordinate System (WCS; \citealt{2002A&A...395.1077C}), where the pixel size is the same as the image cutouts. For each mock galaxy, we pick the object image of the matched LS DR9 galaxy from the image library. Note that since our mock galaxy has very similar redshift and luminosity as the LS source galaxy, we do not need to make additional adjustments. We insert the source galaxy image according to the celestial coordinate of the mock galaxy. After all the galaxy images are inserted, the remaining pixels, which are not been assigned values yet, are considered to be dominated by noise. These pixels are filled with values sampled from the set of background pixels near the source galaxy.

\end{itemize}

Here we do not include the images of stars, which if necessary will be added for specific purposes.
Figure~\ref{fig:imaging} shows the mimic images around halos of different masses at different redshifts, with member galaxies marked by small red circles. As a comparison, the observed images of the DESI LS groups/clusters extracted from the DR9 group catalog are also given. The foreground and the background of the observed images seem more complex. It is due to the cut of magnitude and redshift in the mock catalog. The mimic LS-like image successfully reproduces the diverse galaxy distribution and multi-band morphology within various dark matter halos. 
Such kind of realistic images are instrumental in the more comprehensive modeling of slitless spectroscopic observations, involving overlapping spectra. This will enable us to provide substantial constraints on the selection effects of slitless spectroscopic redshift surveys. 


In terms of the stellar mass of galaxies, there is one interesting feature in our MGRS. As we are using the maximum mass of the subhalos during their evolution pass to assign galaxy luminosities, there are some rare cases $\sim 2000$ galaxies  (comprising approximately 130 million entries) whose stellar mass is larger than the survived subhalo mass.  These objects thus show up as dark matter deficient galaxies. However, it should be noted that this prediction assumes the stellar mass is not stripped after infall according to the abundance matching model, which may not be true for severely stripped subhaloes. 

\section{The Construction of MGRS and the group catalog}\label{sec_geo}

In this section, we construct the MGRSs with realistic survey geometries as well as group catalogs for the CSST and DESI LS observations. 

\subsection{The footprints of CSST and DESI LS} 
The CSST optical survey\footnote{\url{http://www.nao.cas.cn/csst/}} plans to produce an unprecedented joint wide survey of $\sim 17500$ square degrees for both the photometric and slitless spectroscopic surveys. There are seven bands (NUV, $u$, $g$, $r$, $i$, $z$, and $y$) in the photometric imaging survey, and three wide bands (GU, GV, and GI) in the slitless spectroscopic survey \citep[][]{2019ApJ...883..203G, ZhanHu2021...CSB}. 
The footprints of CSST wide surveys are bounded by $|\beta| > 23.43^\circ$ in ecliptic coordinates and $|b| > 15^\circ$ in Galactic coordinates. 

We use the Healpix tool \citep{2005ApJ...622..759G, Zonca2019} to map the footprint of the MGRS. It can divide the spherical surface into subdivisions which each subdivision covers the same surface area as every other subdivision. We set the parameter $\rm nside = 256$, which corresponds to $\rm 5.246\times 10^{-2} deg^2$ per subdivision. The DESI LS DR9 footprint is defined as the summary of the subdivisions which includes any object in the galaxy catalog, as detailed in Section \ref{sec:DESI-obs}. 
This DESI LS DR9 footprint covers approximately $\rm 18350\ deg^2$ of the sky. It's also important to note that regions near the galactic plane with high stellar density, specifically where $|b| < 25^\circ$, have been excluded from the footprint.

In addition to these survey geometry cuts, we also apply the foreground mask to our MGRSs. The foreground sources include globular clusters, planetary nebulae, nearby large galaxies, and Gaia stars with $G < 16$. The detailed parameters of masking geometry are provided by the external catalogs of LS DR9 used for masking\footnote{\url{https://www.legacysurvey.org/dr9/external/##external-catalogs-used-for-masking} }. For foreground large galaxies, a mask with elliptical geometry is applied. 
Overall, an area of 938 square degrees is masked within the footprints. About 5\% of galaxies in the MGRS will be removed by applying the foreground mask. 


Figure~\ref{fig:footprint} gives a brief summary of the geometry of the DESI LS and CSST surveys. Besides what was mentioned above, we did not apply any other effects, especially the redshift completeness, etc. Those effects will be incorporated in the subsequent studies, while this sample serves as the benchmark for future test studies. 

\subsection{Magnitude and redshift cut} \label{sec:cut}
\begin{figure}
\centering
\includegraphics[width=0.5\textwidth]{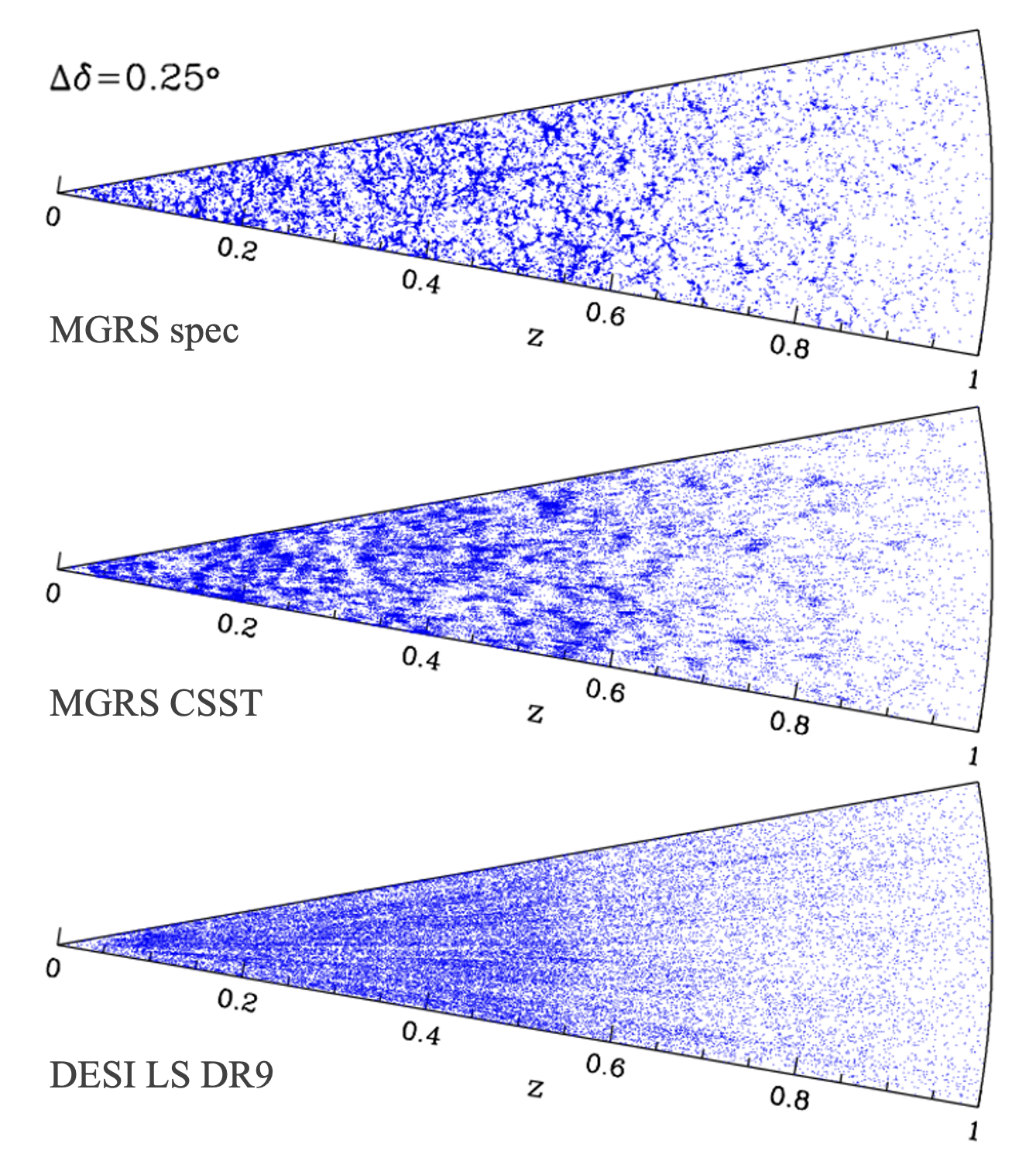} %
\caption{A slice of galaxy distributions in our MGRS using spectroscopic (top) and CSST slitless spectroscopic (middle) redshifts, compared with those in the  DESI LS DR9 with photometric redshifts (bottom).}
\label{fig:zdist} 
\end{figure}

\begin{figure*}
\centering
\includegraphics[width=1.\textwidth]{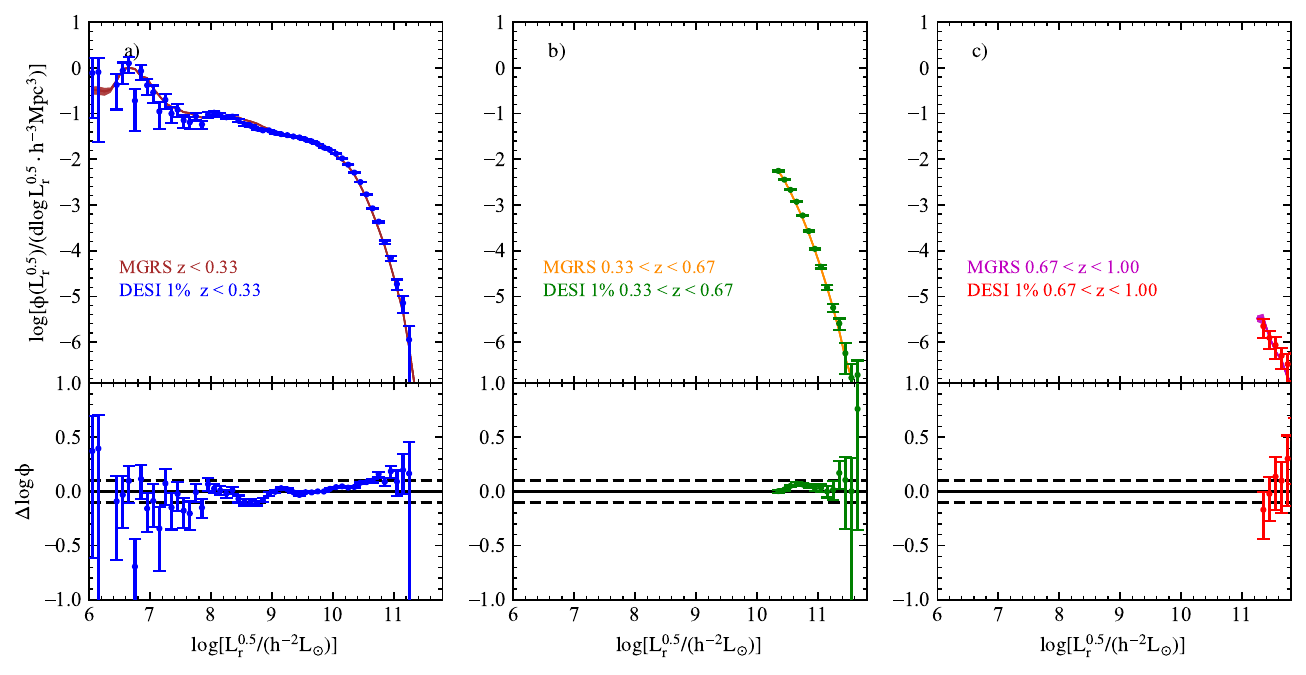}
  \caption{Top: The $r$-band luminosity functions in three redshift bins of MGRS (brown, orange, and magenta) and DESI 1\% (blue, green, and red). The error bars are estimated from the bootstrap method with 100 times resampling.  Bottom: the difference in luminosity functions between the mock catalog and the DESI 1\% observation. The black solid lines ($\Delta = 0$) and dashed lines ($\Delta = \pm 0.1$) serve as reference lines.
  }
  \label{fig:lfr}
\end{figure*} 

\begin{figure*}
\centering
  \includegraphics[width=1.\textwidth]{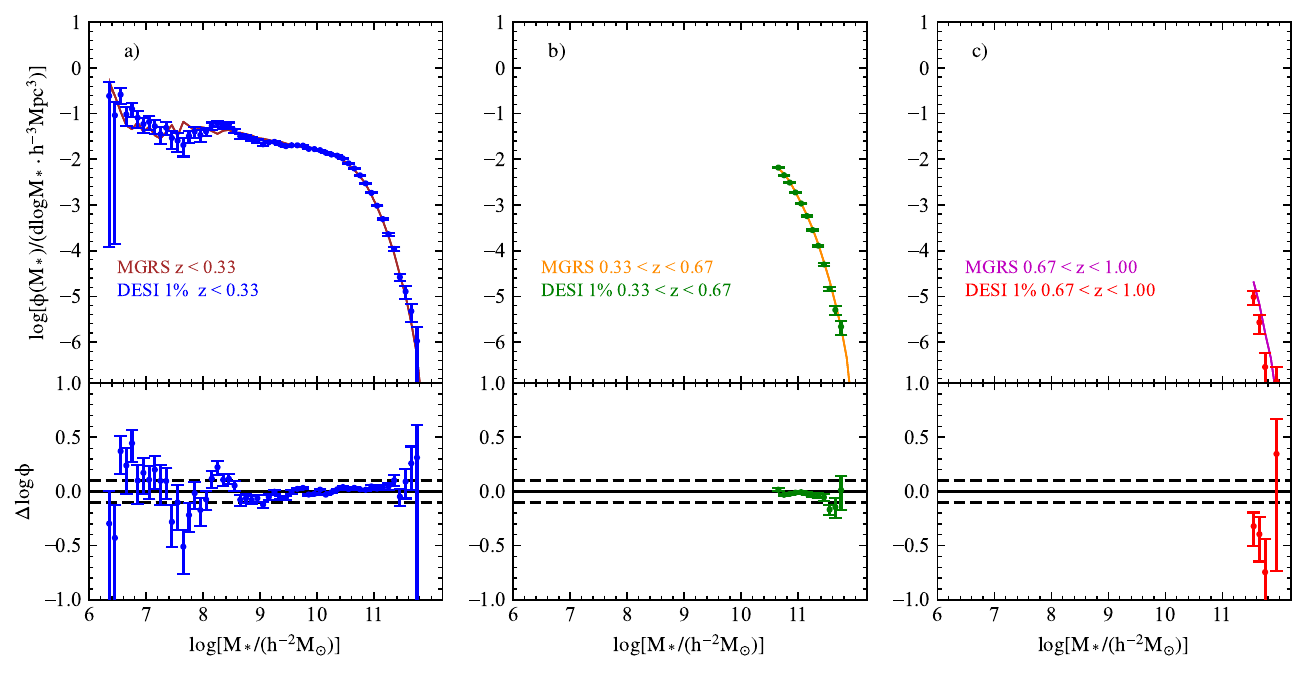}
  \caption{Same as Figure \ref{fig:lfr}, but for the stellar mass functions. Lines, colors, and symbols are the same as those in Figure  \ref{fig:lfr}.   
}
  \label{fig:mf}
\end{figure*}

The vast majority of our mock galaxy properties are sampled from the DESI LS DR9 observation, with a magnitude limit cut $m_z<21.0$, and a redshift cut $z<1.0$. We also apply the same magnitude and redshift cuts to ensure completeness, although the $z$-band imaging of CSST (or DESI LS) is at least 2 mag deeper. 
The sensitivity of reddest band GI is 23.2 mag for point sources in the CSST wide survey \citep{ZhanHu2021...CSB}. 
The GI band is essential for the determination of redshift, especially at $z < 1$. 
Within this redshift range, the majority of emission lines for redshift determination is covered by the GI band of $620-1000 \rm nm$ (e.g., [SII 6730], [NII 6585], [H$\alpha$ 6563], [OIII 4959, 5007] and [OII 3726, 3728]). 
Nevertheless, if the CSST slitless observation turns out to be able to detect fainter galaxies, we can extend this work using those observational data obtained by Hyper Suprime-Cam \citep[HSC;][]{2018PASJ...70S...4A} and Prime Focus Spectrograph \citep[PFS;][]{2014PASJ...66R...1T, 2022arXiv220614908G}.  

In this work, we provide four sets of redshifts for our recent studies, which are specified as follows. 
\begin{itemize}
\item $z_{\rm cos}$, the cosmological redshifts calculated according to the real space distribution of galaxies. 

\item $z_{\rm spec}$, the redshifts of galaxies calculated taking into account the peculiar velocities and a typical redshift error 35 ${\rm km~s^{-1}}$ in the current spectroscopic redshift surveys. 

\item $z_{\rm CSST}$, the redshifts of galaxies calculated by adding a photometric redshift error with $\sigma_z=0.003(1+z_{\rm spec})$, mimicking the CSST slitless spectroscopic redshift error. We will update these once our CSST slitless spectra and redshift emulators are ready \citep[]{2024arXiv240104171W}. 

\item $z_{\rm photo}$, the photometric redshifts of galaxies calculated by adding a photometric redshift error with $\sigma_z=(0.01+0.015z_{\rm spec})(1+z_{\rm spec})$, mimicking the DESI DR9 photometric redshift error.
\end{itemize}

As an illustration, we show in Fig. \ref{fig:zdist} the projected distribution of a small selection of galaxies with the thickness of $\Delta \delta =0.25^\circ$. Shown in the upper, middle, and lower panels are galaxies in the MGRS with $z_{\rm spec}$, in the MGRS with $z_{\rm CSST}$, and in the DESI DR9 with photometric redshifts, respectively. 

\subsection{Finding galaxy groups} 

Dark matter halos are the building blocks of our Universe. Their number density (halo mass function) and space distribution (bias) hold important information about the cosmology \citep[e.g.][]{2022ApJ...936..161W}. Their positions can be used to stack weak lensing signals and provide better scaling relations \citep[e.g.][]{2022MNRAS.511.3548S, 2023MNRAS.523.4909Z}. They also provide the main environment that regulates the formation and evolution of galaxies \citep[e.g.][]{2018ApJ...852...31W}.  A halo-based group finder can directly map the distribution of dark matter halos from a volume or flux-limited galaxy observations \citep[e.g.][]{2007ApJ...671..153Y}. 

In addition, as pointed out in \citet{2021ApJ...909..143Y}, for a photometric redshift survey, the rich groups with members larger than 10 can significantly reduce the redshift error of the system. As the future CSST slitless spectroscopic redshift survey still has quite large redshift uncertainties, galaxy groups will provide a supplementary way to increase the redshift accuracy for cosmological studies.  The feasibility of such kind of probes will be carried out in subsequent studies. 

Here we adopt the same extended halo-based group finder developed in \citet{2021ApJ...909..143Y} to our MGRSs. The group positions, mass, total luminosity, as well as galaxy memberships are provided for all the galaxies in the MGRSs.  In this study, we mainly focused on the group catalogs constructed using the spectroscopic redshifts $z_{\rm spec}$. Other versions of groups will be studied as well.

\begin{figure*}
\centering
  \includegraphics[width=1.\textwidth]{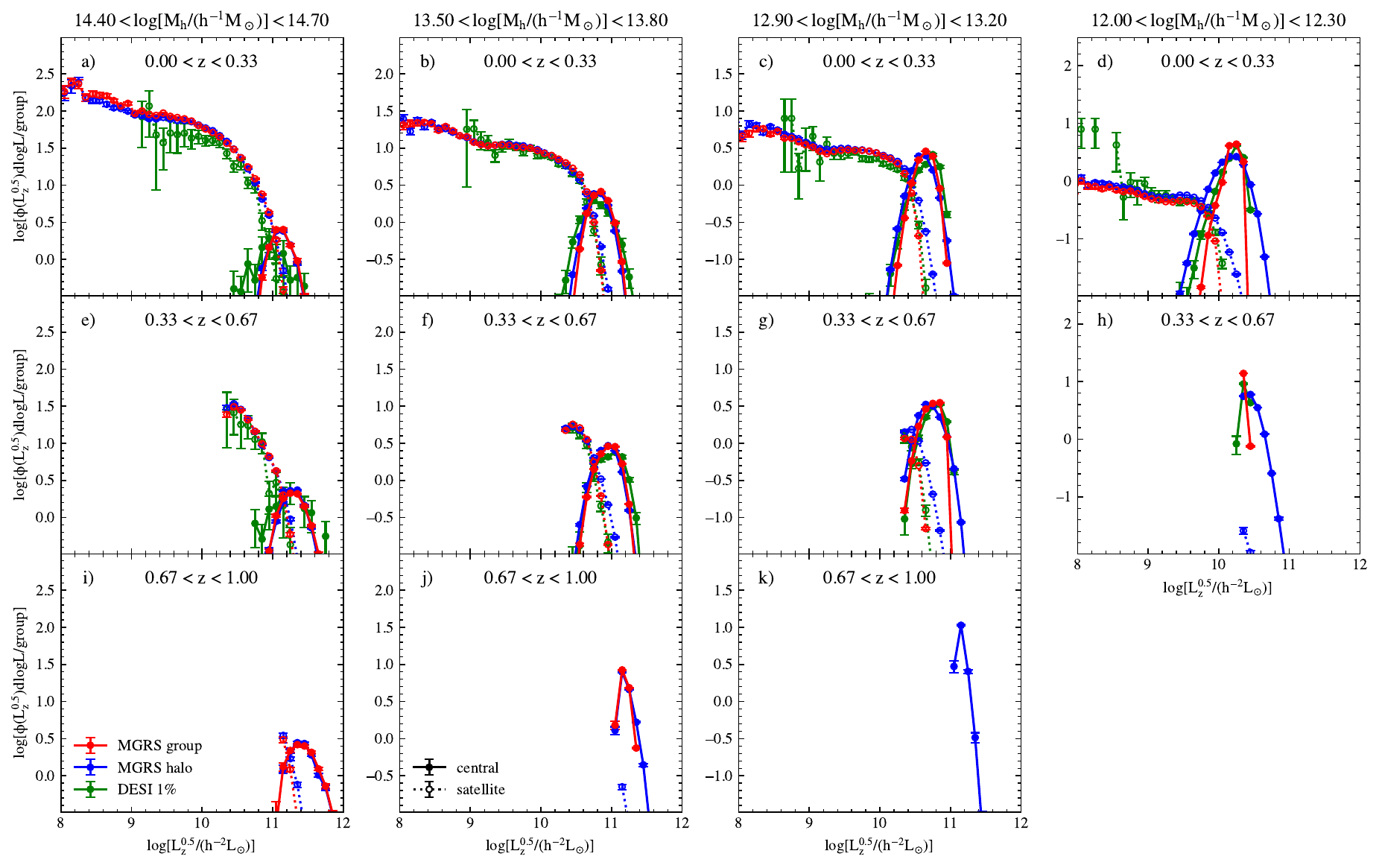}
  \caption{The z-band CLFs of MGRS and DESI 1\% (green) using the magnitude sample with $m_z < 19.0$. The red colors depict the results of MGRS based on the halos identified by the group finder;  The blue colors are the references depicting the results of MGRS based on the halos given by the FOF method. The filled circles with solid lines represent the CLFs of central galaxies. The unfilled circles with dotted lines represent the CLFs of satellite galaxies. The error bars are estimated from the bootstrap method with 100 times resampling. 
  }
  \label{fig:clf-z}
\end{figure*}

\begin{figure*}
\centering
  \includegraphics[width=1.\textwidth]{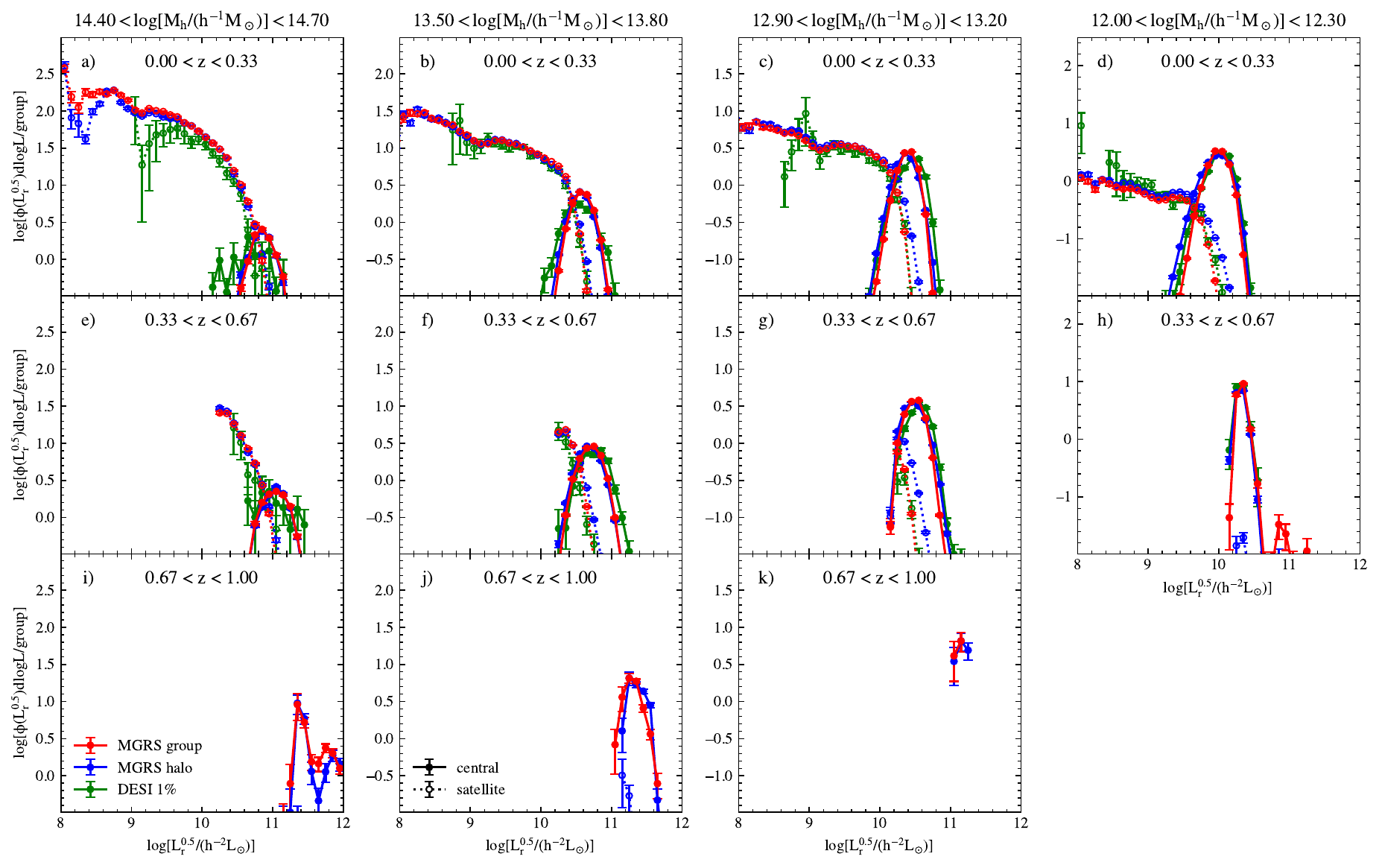}
  \caption{Same as Figure \ref{fig:clf-z}, but for the $r$-band CLF using the magnitude sample with $m_r < 19.5$. Lines, colors, and symbols are the same as those in Figure  \ref{fig:clf-z}. 
  }
  \label{fig:clf-r}
\end{figure*} 

\section{Some basic properties of the MGRS}\label{sec_test}

Since there will be subsequent studies focusing on observational selection effects on various clustering properties, here we only provide some basic properties of our MGRS with the set of spectroscopic redshift in the North Galactic Cap of DESI LS footprint. We compare the statistics of  LFs, MFs, and CLFs between the MGRS and the DESI 1\% observations. The DESI observational measurements are performed in \cite{2312.17459v1}. Readers can refer to that paper for more details. 

\subsection{$r$-band luminosity functions} 
The galaxy luminosity function is one of the most fundamental statistics that quantitatively describes the abundance of galaxies with different luminosity \citep[e.g.,][]{2003ApJ...594..186B}. 
It’s important to note that, by definition, the z-band galaxy LFs of our MGRS should align well with those obtained from DESI 1\% data (refer to Figure \ref{fig:lf}). However, the r-band LFs may deviate due to two potential reasons. Firstly, in our assignment of galaxy properties, we have incorporated all the DESI LS DR9 photometric galaxies with a magnitude of $m_z <21.0$. The vast majority of these galaxies do not have spectroscopic redshifts. Secondly, the color (or other band magnitudes) may not necessarily be accurately predicted in our 3-D parameter space matching process. 


Figure \ref{fig:lfr} shows the comparison of the galaxy $r$-band LFs between the MGRS and the DESI 1\% observation. The $r$-band luminosities of galaxies in MGRS are obtained though the matched DESI galaxy's $r$-band luminosity after $K$-correction (see Section~\ref{sec:properties}).  
Our MGRS can reproduce the observed galaxy luminosity functions within 0.1 dex for galaxies with $\rm L^{0.5}_r > 10^8L_\odot$ and within 1-$\sigma$ level for galaxies with $\rm L^{0.5}_r < 10^8L_\odot$. 
By comparing our measurements with those obtained from the DESI 1\% by \cite{2312.17459v1}, we see a nice agreement between them. 
The overall nice agreement indicates that the $r$-band (as well as other bands) luminosities of galaxies can be fairly well assigned through the 3-D parameter space matching method.

\subsection{Stellar mass functions} 

The stellar mass is a crucial property that characterizes the formation and evolution of galaxies. It is one of the most straightforward parameters that semi-analytic model or hydro-dynamical simulation can predict, and was among the first set of parameters used for model tuning \citep{Katsianis2021, Xie2023}. In our DESI LS DR9 galaxy seed catalog, the stellar mass of a galaxy is determined using the K-correct software \citep{2007AJ....133..734B}. Owing to the procedure that incorporates stellar evolution synthesis based on the Bruzual-Charlot models \citep{2003MNRAS.344.1000B}, this code is capable of providing a preliminary estimation of stellar mass \citep{2007AJ....133..734B}. 

Figure \ref{fig:mf} shows a comparison of galaxy mass functions (MFs) between the MGRS and DESI 1\% observations. By comparing our measurements of the MGRS with those of DESI, we find excellent agreement of MFs for $M_* >10^{8.5}M_\odot$, except in the highest redshift bin where the statistic of DESI 1\% data is poor.  Towards the low mass end, only some minor differences between MGRS predictions and DESI observations appear. 

\subsection{Conditional luminosity functions} 

The CLF $\phi(L|M_h)$ which describes the probability of finding galaxies with luminosity $L$ in a halo with mass $M_h$, is an essential model to link galaxies with dark matter halos \citep[e.g.][]{2003MNRAS.339.1057Y, 2003MNRAS.340..771V}. It can be used to better interpret the clustering of galaxies and hence the constraining of cosmological parameters \citep[e.g.][]{2013MNRAS.430..725V, 2013MNRAS.430..767C}, can be used to evaluate galaxy formation models,  and provides clues about the evolution track of galaxies \citep[e.g.][]{2012ApJ...752...41Y, 2018ARA&A..56..435W}. Apart from the model constraints using the luminosity (stellar mass) functions as well as clustering measurements of galaxies, there have been a number of successful direct measurements from galaxy groups \citep[e.g.][]{2005MNRAS.358..217Y, 2008ApJ...676..248Y, 2009ApJ...695..900Y}. 

The red circles with error bars shown in Figure \ref{fig:clf-z} are the $z$-band CLFs measured from our group catalog constructed from the MGRS, labeled as ``MGRS group''.  Results shown in different panels represent groups within different halo mass ranges, and shown in different rows are for groups within different redshift ranges, as indicated. 
For reference, we also plot the `true' CLFs of the MGRS using blue circles with error bars, labeled as ``MGRS halo'', where halos and subhalos in the Jiutian N-body simulation, as well as their associated galaxies, are used to make the measurements. Overall, the two sets of CLFs agree with each other very well except for some minor discrepancies.  For the central galaxies, the scatter of the CLF measured from groups is slightly underestimated at the lowest halo mass bin, because we are using the ranking of total $z$-band luminosity to estimate the halo mass. 
The good agreement between them, especially the overall amplitude and shape, thus ensures us that the CLFs measured from the group catalog constructed by \citet{2021ApJ...909..143Y} can be used to constrain galaxy formation processes. 

As a comparison, we also show in Figure \ref{fig:clf-z} using green circles the CLFs measured from the DESI 1\% observations for the galaxies.  Due to the insufficient sample size in DESI 1\%, which has relatively small sky coverage, we do not provide the results of 1\% at $0.67 < z < 1.0$.  Although our fiducial SHAM MGRSs are not tuned to recover the subtle galaxy-halo connections, it can be seen that the CLFs in DESI 1\% have a good agreement with the MGRS results. Our MGRSs only have an overproduce  ($\sim $ 0.2 dex) the CLF for satellite galaxies for the massive halo, a fact that can guide us to obtain a better galaxy-halo connection modeling. 


Apart from the assigned $z$-band luminosity, we also investigate the situation for the  $r$-band luminosity of galaxies which is obtained by using a 3-D parameter space matching with the DESI DR9 galaxies.  Figure \ref{fig:clf-r} shows the comparison of the $r$-band CLFs with the DESI BGS  magnitude cut $r<19.5$. 
The results of the $r$-band CLFs are very similar to those of $z$-band as shown in Figure \ref{fig:clf-z}, again showing nice agreement between the measured and true CLFs.   These comparisons
demonstrate that the $r$-band CLFs can also be well recovered, even the galaxy groups are obtained using the $z$-band luminosity of galaxies. The overall similar $r$-band and $z$-band performances of the MGRS also indicate that the 3-D parameter ($z|\log L_z|\log M_h$) nearest neighbor matching has a good performance in determining the galaxy properties. 
The discrepancy between MGRS and DESI 1\% is roughly similar to the $z$-band and slightly larger in the most massive bin. In general, the discrepancy indicates that the standard SHAM we are using may over-predict the satellite population, which might suffer from somewhat more disruption effects. We will carry out a subsequent study on the CLF modeling based on the DESI BGS year 1 dataset, which shall provide a better galaxy-halo connection modeling than the standard SHAM we are currently using.  

\section{Conclusions} \label{sec_conclusion}

In this study, by employing the light-cone catalog of halo/subhalo constructed from one of the state-of-the-art large Jiutian N-body simulations, we present a set of mock galaxy redshift surveys with redshift $0<z<1$ and a magnitude limit of $m_z<21$ for CSST slitless spectroscopic redshift survey evaluations. The main advantages of our MGRSs are summarized as follows.

\begin{itemize}

\item The HBT+ code that we utilized is capable of tracking the evolution of all the subhalos, even if they are disrupted. As a result, our subhalo catalog suffers less from the impact of simulation resolution. 

\item We measured the $z$-band LFs of galaxies at different redshifts directly from the latest DESI 1\% data release, which has very high spectroscopic completeness. These measurements were then utilized to model our mock galaxies. 

\item We implemented a three-parameter ($z|\log L_z|\log M_h$) space sampling method based on the DESI LS DR9. This methodology allows us to assign additional properties to galaxies, thereby ensuring maximum consistency with the observations. 

\item Apart from galaxy properties, we utilize multi-band images for each galaxy. This approach allows us to generate more realistic observational images.

\item In addition to the galaxy catalogs, we have also assembled group catalogs. These hold significant potential for future studies related to cosmology and galaxy formation. 

\item By comparing our MGRS with DESI 1\% observational data, in terms of LFs, SMFs, and CLFs we demonstrate that our fiducial MGRSs based on the standard SHAM already work fairly well. 

\item We have prepared the survey geometry for the CSST and LS DR9 observations using Healpix, along with the associated random galaxy catalog. The foreground masks are also taken into consideration. These are now ready for use in subsequent probes. 
\end{itemize}

Currently, within our MGRSs, we have provided four sets of redshifts ($z_{\rm cos}$, $z_{\rm spec}$, $z_{\rm CSST}$, and $z_{\rm photo}$) with Gaussian errors. The more realist selection effects induced by the slitless spectra, especially morphology self-blending and galaxy-galaxy inter-blending, would be further added using an Emulator developed for CSST Slitless Spectroscopic Redshift Survey \citep{2024arXiv240104171W}. Based on that set of redshifts, subsequent evaluations will be conducted to understand how the selection effects can affect our probe of cosmology and galaxy formation.

\section*{Acknowledgements}

This work is supported by the National Key R\&D Program of China (2023YFA1607800, 2023YFA1607804),  National Science Foundation of China (Nos. 11833005, 11890692, 11973032, 11890691, 11621303, 12273088), “the Fundamental Research Funds for the Central Universities”, 111 project No. B20019, and Shanghai Natural Science Foundation, grant No.19ZR1466800. We acknowledge the science research grants from the China Manned Space Project with Nos. CMS-CSST-2021-A02 and CMS-CSST-2021-A03. We thank the sponsorship from the Yangyang Development Fund. Y.Z.G. acknowledges the support from China Postdoctoral Science Foundation (2020M681281) and Shanghai Post-doctoral Excellence Program (2020218). The computations in this paper were run on the Gravity Supercomputer at Shanghai Jiao Tong University. 

The Photometric Redshifts for the Legacy Surveys (PRLS) catalog used in this paper was produced thanks to funding from the U.S. Department of Energy Office of Science, Office of High Energy Physics via grant DE-SC0007914.

This research used data obtained with the Dark Energy Spectroscopic Instrument (DESI). DESI construction and operations is managed by the Lawrence Berkeley National Laboratory. This material is based upon work supported by the U.S. Department of Energy, Office of Science, Office of High-Energy Physics, under Contract No. DE–AC02–05CH11231, and by the National Energy Research Scientific Computing Center, a DOE Office of Science User Facility under the same contract. Additional support for DESI was provided by the U.S. National Science Foundation (NSF), Division of Astronomical Sciences under Contract No. AST-0950945 to the NSF’s National Optical-Infrared Astronomy Research Laboratory; the Science and Technology Facilities Council of the United Kingdom; the Gordon and Betty Moore Foundation; the Heising-Simons Foundation; the French Alternative Energies and Atomic Energy Commission (CEA); the National Council of Science and Technology of Mexico (CONACYT); the Ministry of Science and Innovation of Spain (MICINN), and by the DESI Member Institutions: www.desi.lbl.gov/collaborating-institutions. The DESI collaboration is honored to be permitted to conduct scientific research on Iolkam Du’ag (Kitt Peak), a mountain with particular significance to the Tohono O’odham Nation. Any opinions, findings, and conclusions or recommendations expressed in this material are those of the author(s) and do not necessarily reflect the views of the U.S. National Science Foundation, the U.S. Department of Energy, or any of the listed funding agencies. 

\section*{Data Availability}

The mock galaxy catalogs for the CSST surveys constructed from Jiutian simulation in this paper are shared through \url{https://gax.sjtu.edu.cn/data/CSST/CSST.html}. 




\bibliographystyle{mnras}
\bibliography{ms} 






\bsp	
\label{lastpage}
\end{document}